\documentclass{aa}  
\usepackage{xcolor}
\usepackage{graphicx}
\usepackage{xcolor}
\usepackage{graphicx}
\usepackage{txfonts}
\usepackage{subfig}
\begin{document}

   \title{Chemical evolution of the Galactic bulge with different stellar populations}

%   \subtitle{I. Overviewing the $\kappa$-mechanism}

   \author{M. Molero
          \inst{1,2,3}\fnmsep%\thanks{Just to show the usage
          %of the elements in the author field}
          \and
          F. Matteucci\inst{2,3}
          \and
          E. Spitoni\inst{3}
          \and
          A. Rojas-Arriagada\inst{4,5,6,7}
          \and
          R. M. Rich\inst{8}
          }

   \institute{Institut f\"ur Kernphysik, Technische Universit\"at Darmstadt, Schlossgartenstr. 2, Darmstadt 64289, Germany\\
              \email{mmolero@theorie.ikp.physik.tu-darmstadt.de}
         \and
             Dipartimento di Fisica, Sezione di Astronomia, Università degli studi di Trieste, Via G.B. Tiepolo 11, I-34143 Trieste, Italy
        \and
            INAF, Osservatorio Astronomico di Trieste, Via Tiepolo 11, I-34131 Tireste, Italy
        \and
            Departamento de F\'isica, Universidad de Santiago de Chile, Av. Victor Jara 3659, Santiago, Chile
        \and
            Millennium Institute of Astrophysics, Av. Vicu\~na Mackenna 4860, 82-0436 Macul, Santiago, Chile 
        \and
            Center for Interdisciplinary Research in Astrophysics and Space Exploration (CIRAS), Universidad de Santiago de Chile, Santiago, Chile
        \and
            N\'ucleo Milenio ERIS
        \and
            Department of Physics and Astronomy, UCLA, 430 Portola Plaza, Box 951547, Los Angeles, CA 90095-1547, USA
             }

   \date{Received September 15, 1996; accepted March 16, 1997}

% \abstract{}{}{}{}{} 
% 5 {} token are mandatory
 
  \abstract
   {The metallicity distribution function (MDF) of the Galactic bulge is characterized by a multi-peak shape, with a metal-poor peak centered at $\mathrm{[Fe/H]\sim-0.3}$ dex and a metal-rich peak centered at $\mathrm{[Fe/H]\sim+0.3}$ dex. The bimodality of the MDF is also reflected in the [$\alpha$/Fe] versus [Fe/H] abundance ratios, suggesting the presence of different stellar populations in the bulge.}
   {In this work we aim to reproduce the observed MDF of the Galactic bulge by testing a scenario in which the metal-poor component of the bulge is formed by stars formed in situ, during a strong burst of star formation, while the metal-rich population is formed by stars created in situ  during a second burst of star formation and/or stars accreted from the innermost part of the Galactic disk as an effect of a growing bar.}
   {We adopted a chemical evolution model that is able to follow the evolution of several chemical species with detailed nucleosynthesis prescriptions. In particular, because of the importance of the production of Fe in constraining the MDF, close attention is paid to the  production of this element in both Type Ia supernovae and massive stars. In particular,  we included yields from rotating massive stars with different rotational velocity prescriptions. Our model also takes the infall and outflow of gas into account, as well as the effect of stellar migration. Results are compared to $\sim$13000 stars from the SDSS/APOGEE survey that belong to the region located at a
Galactocentric distance $\mathrm{R_{GC}\leq3.5\ kpc}$.}
   {We successfully reproduce the observed double-peak shape of the bulge MDF as well as the abundance trends of the $\alpha$ elements relative to Fe by assuming both (i) a  multi-burst star formation history with a quenching of the first burst of $\mathrm{\sim 10^2\ Myr}$  and (ii) migration of stars from the innermost part of the Milky Way disk, as an effect of a growing bar. According to our results, the fraction of the stellar mass of the bulge-bar that belongs to the inner disk is $\mathrm{\sim40\%}$. In terms of the nucleosynthesis, we conclude that models that assume either no rotation for massive stars or a distribution of rotational velocities that favors slow rotation at high metallicities best reproduce the observed MDF as well as the [$\alpha$/Fe] and the [Ce/Fe] versus [Fe/H] abundance patterns.}
   {}


   \maketitle
%
%-------------------------------------------------------------------

\section{Introduction}
\label{sec: Introduction}

The formation and evolution of the Milky Way (MW) bulge has been the subject of intense studies over the last decade. According to the original picture, the Galactic bulge can be considered a classical bulge, namely a spheroidal remnant of mergers of primordial structures in a Lambda cold dark matter context ($\Lambda$CDM) (\citealp{Ortolani1995, Baugh1996, Abadi2003, Abadi2003_2}). However, both observations of our own bulge (e.g., \citealp{Binney1991, Bissantz2002, Lopez2005}) and at higher redshifts (e.g., \citealp{Tacchella2015, Nelson2016, Shen+23}) point toward a much more complex picture. Contrary to classical bulges, pseudo-bulges should form from disk stars due to the vertical instability of a stellar bar (\citealp{combes1990}). This scenario should lead to the formation of a triaxial boxy bar structure, the so-called boxy/peanut (B/P) shape of the Galactic bulge (e.g., \citealp{Weiland1994, McwilliamZoccali2010, wegg2013, Ness2016}), which in $N$-body simulations has also been seen to form from barred stellar disk galaxies (\citealp{Raha1991, Oneil2003, Athanassoula2005, Shen2010, Ciambur1021, Gosh2023}). Moreover, fully formed bulges or a high central star formation (SF) are observed in galaxies at redshift $\mathrm{z\sim2}$(\citealp{Tadaki2017}); this suggests that those bulges assembled before the formation of the bar, which, in the case of MW-type galaxies, happened at $\mathrm{z\sim1}$ (i.e.,$\mathrm{\sim8\ Gyr}$ ago). Therefore, the mechanism responsible for the creation of such structures should be a fast and strong primordial collapse of gas responsible for at least one rapid SF episode. 

The Galactic bulge region can be very hard to observe because of heavy extinction and crowding. Nevertheless, several spectroscopic (e.g., BRAVA, \citealp{Rich2008, Kunder2012}, \textit{Gaia}-ESO, \citealp{gilmore2012}, APOGEE,  \citealp{Majewski2017}, ARGOS, \citealp{freeman2013}, GIBS, \citealp{Zoccali2014}) and photometric (e.g., the VVVX survey; \citealp{Minniti2010}) surveys have been developed to shed light on the history of the bulge (see \citealp{chiappini2018} for a review of the different surveys). The picture that emerges from observations appears, however, to be complex. The metallicity distribution function (MDF) of stars observed in the bulge region has a bimodal shape (observed for the first time by \citealp{Hill2011} and later confirmed in a large number of studies, e.g., \citealp{bensby2011, uttenthaler2012, gonzalez2015, rojas2017, zoccali2017, Rojas-Arriagada2019, Rojas-Arriagada2020, Queiroz2020, Queiroz2021, Johnson2020, johnson}), which can be an indication of two (or more) stellar populations: a metal-rich (MR) population centered at $\mathrm{[Fe/H]\sim 0.25}$ dex and a metal-poor (MP) population centered at $\mathrm{[Fe/H]\sim -0.3}$ dex. The two stellar populations can also have different kinematics, with the MR one rapidly rotating and dynamically cold and the MP one dynamically hotter and rotating more slowly. \citet{Babusiaux2010}, who inspected the kinematics of the large sample of over 500 red-giant branch (RGB) stars presented in \citet{zoccali2008}, concluded that the MR population shows a vertex deviation compatible with the MW bar (it is worth noting that vertex deviation was first seen by \citealp{Zhao1996}  and \citealp{ soto2007}), while the MP population is compatible with a spheroid (and/or a thick disk). Moreover, results of the chemodynamical model presented by \citet{Portail2017} show that MR stars (with $\mathrm{[Fe/H]\geq-0.5}$ dex) are strongly barred with dynamical properties consistent with a common disk origin, while MP stars (with $\mathrm{[Fe/H]<-0.5}$ dex) show more kinematic variations with metallicity, which is interpreted as being due to the contributions from different stellar populations.

As pointed out by \citet{baba2020}, if the Galactic bar significantly impacts the dynamic of the stars in the bulge (and therefore also in the inner disk; e.g., \citealp{Minchev2016}), identifying its formation epoch should provide important insights into the history of the MW. As discussed above, the time of the formation of the Galactic bar is highly uncertain. From the distribution of infrared carbon stars, \citet{cole2002} estimated an age of $\mathrm{\sim2\ Gyr}$. From observations of luminous face-on spiral galaxies, \citet{sheth2008} analyzed the variation with redshift of the fraction of galactic bars and estimated that the bar in spirals with masses similar  to that of the MW should form at $\mathrm{z\simeq1}$ ($\mathrm{\sim 8\ Gyr}$ ago; a result that was confirmed by the zoom-in cosmological simulations in \citealp{Kraljic2012}). More recently, the bar formation and building epoch were also estimated to have happened $\mathrm{8-9\ Gyr}$ ago from a study of the \textit{Gaia} Data Release (DR) 2 set of long-period variables by \citet{Grady2020}, though the long-period variables in the bar are generally more MR and hence may be older at a given luminosity.  It should be pointed out that the age of the stars in the bar does not necessarily correspond to the age of the bar itself, since the bar can capture stars that formed before its formation (though the absence of SF in the bar cannot be excluded a priori; see e.g., \citealp{Anderson2020}). This process is described by \citet{chiba2021} (see also \citealp{Chiba2021b, chiba2022}). According to the authors, the bar experiences angular momentum loss due to dynamical friction from the dark matter halo, which slows its pattern speed, $\Omega_{\rm p}$ (see also \citealp{Hernquist1992, debattista2000, martinez-valpuesta2006}). When the bar slows down, the resonance sweeps radially outward in radius throughout the disk, sequentially capturing and dragging new stars. A fraction of stars that are trapped in the corotation radius can then be captured by the bar itself. So, from a chemical point of view, stars in the bar should reflect the composition of the location where they where trapped (assuming it is also where they were born). As a consequence, if the bar captures stars from the disk, those stars may have the same composition and age as the disk.

The first chemical evolution model for the Galactic bulge that released the instantaneous recycling approximation was developed by \citet{matteuccibrocato} to explain the results of \citet{Rich1988} and \citet{rich1990}. \citet{matteuccibrocato} suggested that the bulge formed  during a strong burst of SF on a short timescale ($\mathrm{\sim0.5\ Gyr}$), with a more top-heavy initial mass function (IMF) than that of the solar neighborhood (e.g., \citealp{Scalo1986}). Their model predicted a plateau in the [$\alpha$/Fe] ratios in bulge stars longer than that in the solar vicinity, which was later confirmed by \citet{Mcwilliam1994} and is now an observationally established fact. The prescriptions of the model were then confirmed in updated versions (e.g., \citealp{Ballero2007, cescutti2011}). Later on, more chemical evolution models started to try modeling the bimodal MDF of the bulge (e.g., \citealp{GriecoBulge2012, Tsukimoto2012, Grieco2015}) in an attempt to explain it as due to a second infall and/or accretion episode. More recently, \citet{Matteucci2019} successfully reproduced the MR peak of the bulge MDF by assuming a pause of $\mathrm{\sim250\ Myr}$ in the SF of the bulge and proposed a scenario in which the MR population is made of  either stars formed in a second burst after the pause or stars formed in the inner disk and transported into the bulge during the early secular evolution of the bar. 

The main goal of this work is to confirm (or disprove) this hypothesis by adopting a chemical evolution model that includes both multiple SF episodes and the accretion of stars from the MW disk. The bulge has a complex structure with a vertical abundance gradient (e.g., \citealp{zoccali2008, zoccali2017, ness2013, johnson2014, Johnson2020, johnson}). Our models do not address this vertical gradient, but they may shed light on some of the physical processes that could cause a vertical abundance gradient.

The paper is organized as follows: In Section 2 we describe the chemical evolution models and their prescriptions. In Section 3 we present the results regarding the hypothesis that bulge stars formed only in situ, while in Section 4 we present the results for accreted stars. Finally, in Section 5 we draw some conclusions.

\section{Observational data}

We compared our model predictions to the APOGEE DR16 sample of bulge stars of \citet{Rojas-Arriagada2020}. We refer the reader to that paper for the full details concerning the description of APOGEE survey data and the sample selection. We provide a brief summary here.
APOGEE is an SDSS-III and SDSS-IV high-resolution near-infrared spectroscopic survey of resolved stellar populations in the MW and some of its satellites. The main targets of the survey are giant stars, RGB, red clump, and asymptotic giant branch (AGB) stars, which allow large spatial regions to be mapped. The targets were observed in the H-band using two twin high resolution ($R\sim22500$) spectrographs operating at the Apache Point Observatory's 2.5 m Sloan Telescope and Las Campanas Observatory's 2.5 Ir\'en\'ee du Pont telescope. APOGEE spectra are extracted, wavelength calibrated and RV corrected using a custom pipeline \citep{Nidever2015}. Stellar atmospheric parameters and abundances of up to 26 elements are measured using the APOGEE Stellar Parameters and Chemical Abundances Pipeline (ASPCAP; \citealp{Garcia2016}).

In \citet{Rojas-Arriagada2020} an initial sample was selected within the region $|l|<16^\circ$ and $|b|<15^\circ$. The individual fields are not homogeneously distributed over that region but are concentrated toward the midplane, where extinction levels are high and the near-infrared capabilities of APOGEE can sample bulge stellar populations through the dust veil. A set of spectrophotometric distances were computed using an isochrone fitting technique. In order to secure reliable ASPCAP parameters, several cuts based on rather technical flags were implemented. An additional cut on surface gravity, removing stars with log(g)>2.2 dex, was adopted to ensure a sample free of selection and analysis bias. Finally, from the analysis of several kinematic and orbital parameters, a cut on Galactocentric distance of $R_{GC}<3.5$~kpc was adopted to select a sample of stars spatially located in the bulge region. The bulge sample so defined consists on 13031 stars.

We emphasize that while only metallicity was used to study the shape of the bulge MDF in \citet{Rojas-Arriagada2020}, the complete assortment of abundances is available for the selected sample. On the other hand, although there is a new public APOGEE data release (DR17) available, we adopted the values from DR16 to for the sake homogeneity and consistence with the definition of the sample, which is based on the parameters from the DR16. In the following, we compare our model predictions with both the MDF and the abundances ratios of the $\alpha$ elements Mg, O, Al, and Si as well as with the neutron capture element Ce. We note that this model does not attempt to consider any spatial variation in the \citet{Rojas-Arriagada2020} abundance distribution. 

\section{The model}
\label{sec: the model}

The chemical evolution model adopted for the Galactic bulge is similar to the one developed by \citet{grieco} and later updated by \citet[see also \citealp{Ballero2007, Cescutti&Matteucci2011}]{Matteucci2019}. We assumed that the bulge forms by fast infall of gas with a timescale $\mathrm{\tau_b=0.1\ Gyr}$. The gas is very efficiently converted into stars with a SF parametrized by a Schmidt-Kennicutt law (\citealp{Schmidt1959, Kennicutt1998}) with exponent $k=1.4$ and with an efficiency of SF $\nu=25\ \rm Gyr^{-1}$, much higher than what assumed in the thin disk. The adopted IMF is a \citet{Chabrier2003}, though we investigated the possibility of using a \citet{Salpeter1955} IMF, as done in the previous chemical evolution works. 

After the first main SF episode, mainly responsible for the formation of the MP population observed in the MDF, a second SF burst is present with a delay with respect to the first of order $\mathrm{10^{2}\ Myr}$. The second burst is partially responsible for the creation of the MR population but, as we show in the next sections, it is not enough to completely populate the MR peak in the MDF. Here, we assumed that a fraction of stars originally belonging to the innermost part of the MW disk also populates the Galactic bulge and is partially responsible for the formation of the MR peak. The disk is modeled as described in \citet{Molero2023}, by means of the revised two-infall model of \citet[see also \citealp{Spitoni2019, Spitoni2021}]{Palla2020}. The model assumes that the MW disk forms as a result of two distinct accretion episodes of gas. The first one, with a timescale of $\mathrm{\tau_{d,1}=1\ Gyr}$, is responsible for the formation of the chemically thick disk (namely the high-$\alpha$ sequence observed in the [$\alpha$/Fe] versus [Fe/H] plane), while the second infall episode, delayed with respect to the first one by $\mathrm{\sim3.25\ Gyr}$, is responsible for the formation of the chemically thin disk (the low-$\alpha$ sequence). The timescale for the second infall episode is a function of the Galactocentric distance (see \citealp{chiappini2001}):
\begin{equation}
    \tau_{d,2}(R)=\Big(1.033\frac{R}{kpc}-1.267\Big)\ Gyr,
\end{equation}
so it is shorter toward inner regions.

For both the Galactic bulge and the inner disk the same nucleosynthesis prescriptions are adopted, similar to those adopted by \citet{Molero2023}. We refer to that paper for a more detailed discussion. 

In summary, we used:
\begin{itemize}
    \item The non-rotational set of yields of the FRUITY database (\citealp{Cristallo2009, Cristallo2011, Cristallo2015}) for low- and intermediate-mass stars (LIMSs; $\mathrm{M\leq8\ M_\odot}$).
    \item The \citet{Limongi2018} recommended yield set R with mass-loss and rotation for massive stars ($\mathrm{M\geq13\ M_\odot}$). Here, we also assumed that 20\% of massive stars with $\mathrm{10-25\ M_\odot}$ explode as magneto-rotational supernovae (SNe), with yields from the \citet{Nishimura2017} model L0.75.
    \item Yields from the \citet{Iwamoto1999} model W7 for Type Ia SNe.
    \item For merging neutron stars (MNSs), we adopted yields scaled to those of Sr from the kilonova AT2017gfo measured by \citet[see \citealp{Molero2021}]{Watson2019}. MNSs are assumed to merge with a delay time distribution function (see \citealp{Simonetti2019}).   
\end{itemize}

\begin{figure}
    \centering
    \includegraphics[width=1\columnwidth]{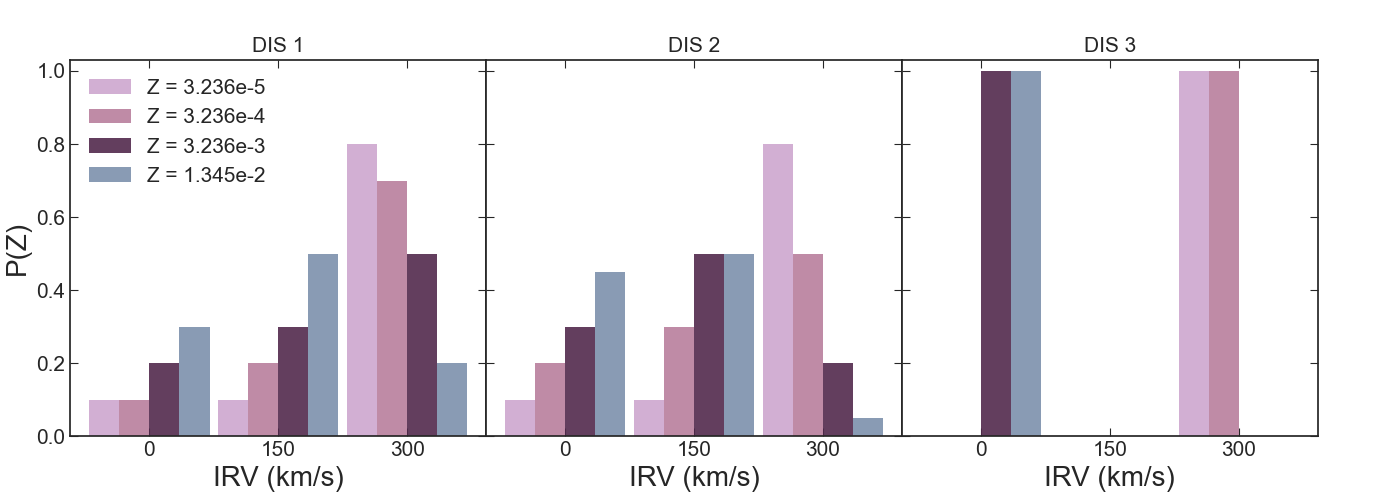}
%    \captionsetup{width=1\textwidth}
    \caption{Probability distributions of initial rotational velocities (IRV) for massive stars as a function of the metallicity (Z).}
    \label{fig: distributions}
\end{figure}

In particular, for massive stars, besides testing the three sets of yields of \citet{Limongi2018} corresponding to three initial rotational velocities (0, 150, and 300 $\rm km/s$), three more sets have been considered, corresponding to three different theoretical distributions for the initial rotational velocities. These new sets of yields were obtained by assuming that the probability that a star rotates at a certain speed is a function of the metallicity, $Z$, with faster initial rotational velocities being most likely at lower $Z$. In this way, we obtained the first two distributions (DIS 1 and DIS 2) reported in Figure \ref{fig: distributions}. In DIS 2, velocities of $0\ \rm km/s$ and of $150\ \rm km/s$ are slightly more likely than in DIS 1 at intermediate and high metallicities. The third distribution shown in the figure, DIS 3, is the one adopted by \citet{romano2019} for studying the CNO isotopes, according to which rotation becomes negligible beyond a metallicity threshold equal to $Z=3.236\times10^{-3}$ (corresponding to [Fe/H]=-1 dex). All the three distributions are supported by the fact that massive stars are expected to rotate faster at lower metallicities, where they are more compact (see \citealp{Klencki2020}). This view is supported both theoretically (\citealp{Frischknecht2016}) and by the observations of an increase ratio of Be/B-type star with increasing metallicity (\citealp{Martayan2007, Martayan2007b}), as well as by the presence of faster rotating massive stars in the Small Magellanic Cloud than in the MW (\citealp{Hunter2008}) and by observations of stars in the globular cluster NGC6522 with high abundances of s-process elements (\citealp{Chiappini2011}). We assumed a flat distribution of the initial rotational velocities with the stellar mass. Although this is a simplification, it still finds agreement in theoretical computations (e.g., \citealp{Frischknecht2016}), which show that, for the same metallicity and initial ratio of surface velocity to critical velocity ($v_{\rm ini}/v_{\rm crit}$), the changes in the surface velocity during the MS phase as a function of mass are relatively small and in any case less than the variations as a function of metallicity.
\\
\\
\subsection{The Galactic bar}
As mentioned in section \ref{sec: Introduction}, the MW bar does not exhibit static rotation. Indeed, simulations of the Galactic bar in the presence of dark matter indicate that the bar experiences angular momentum loss, leading to a decrease in its rotational frequency and an expansion of the bar (\citealp{Hernquist1992, Martinez2006, Bhattarai2022}). As described by \citet{chiba2021}, as the bar decelerates, the resonance regions sweep through the stellar phase-space, capturing and dragging a number of stars. \citet{Chiba2021b} estimated an increase in corotation radius, $\mathrm{R_{cr}}$, since the formation of the bar (here assumed to have happened 8 Gyr ago) of at least $\Delta\mathrm{R_{cr}=1.6\ kpc}$. If we assume that the bar also grows by the same fraction, then\begin{equation}
    \Delta R_b = \Delta R_{cr} \frac{R_b}{R_{cr}} \simeq 1.2\ kpc,
\end{equation}
where $R_{\rm b}=5\ \rm kpc$ is the bar half-length (\citealp{Wegg2015}) and $R_{\rm cr}=6.6\ \rm kpc$ (\citealp{Chiba2021b, Clarke2022}). The fraction of stellar mass swept away by the growing bar is then\begin{equation}
    f=\frac{2\pi\Delta R_b \times R_b \Sigma (R)}{M_b},
\end{equation}
where $M_{\rm b}\simeq10^{10}\ \rm M_\odot$ is the bar mass (\citealp{2016Bland-Hawthorn}). Therefore, given a stellar surface mass density predicted by our model at $R_{\rm GC} \simeq 5\ \rm kpc$ of $\Sigma(5\ \rm kpc) \simeq 143\ \rm M_\odot/pc^{-2}$, we can compute the fraction of stellar mass that is trapped by the growing bar at $\mathrm{\sim 5\ kpc}$, which is $\mathrm{f\simeq53\%}$. This must be considered an upper limit since (i) not all the stars that are trapped into the corotation resonance will also be part of the bar and (ii) only a fraction of stars trapped in the bar will live long enough to populate the Galactic bulge region (which in our model extends until 3 kpc).

From the observational and kinematical point of view, stars that were originally in a disk orbit, will end up in a bar orbit and will then be observed preferentially close to the mid-plane, kinematically colder than MP stars and characterized by ordered orbits. This interpretation is supported by the kinematics of the metal-richest stars in our sample, which we recall is investigated in \citet{Rojas-Arriagada2020}. The kinematic of the studied sample agrees well with both the cylindrical rotation pattern predicted by N-body simulations of bulges formed by secular evolution of the disk through the influence of a forming bar and with observations of external local galaxies with a bar-dominated morphology. The rotation pattern in the sample is also observed to be enhanced toward the mid-plane, which has been interpreted as a signature of a bar-dominated dynamics (see \citealp{Zhao1996, Molaeinezhad2016, Gomez2018}). For a detail discussion about the kinematic of the adopted sample we refer to \citet{Rojas-Arriagada2020}.

\section{Results without accreted stars}

In the following sections, we present results for the MDF and the abundance patterns as predicted by models similar to the best one proposed by \citet{Matteucci2019}, but with updated yields from rotating massive stars, alongside with results obtained from the newly developed models presented in the last Section (see Table \ref{tab: models_bulge}). In this kind of model we assume that the two stellar populations in the bulge (MP and MR) form because of a hiatus in the SF. In Table \ref{tab: models_bulge}, we report the different models tested in this work. In  particular, in the first column we report the name of the model, in the second column the adopted prescriptions for the initial rotational velocity (IRV) of massive stars,  in the third column the duration of the quenching in the SF and in the last column the assumed IMF.

\begin{table}
\centering
\caption{Input parameters for the chemical evolution models. 
}
\label{tab: models_bulge}
\begin{tabular}{lcccc}
\hline
   Model &  IRV  &   $\mathrm{\delta t_{SF}\ (Myr)}$  &   IMF  \\
\hline
model 1 & 0 km/s - constant & 250 & Salpeter \\
model 2 & 150 km/s - constant & 250 & Salpeter \\
model 3 & 300 km/s - constant & 250 & Salpeter \\
\hline
model 4 & 0 km/s - constant & 250 & Chabrier \\
model 5 & 150 km/s - constant & 250 & Chabrier \\
model 6 & 300 km/s - constant & 250 & Chabrier \\
model 7 & DIS 1 & 250 & Chabrier \\
model 8 & DIS 2 & 250 & Chabrier \\
model 9 & DIS 3 & 250 & Chabrier \\
\hline
model 4a & 0 km/s - constant & 0 & Chabrier \\
model 4b & 0 km/s - constant & 150 & Chabrier \\
model 4c & 0 km/s - constant & 350 & Chabrier \\
model 8a & DIS 2 & 0 & Chabrier \\
model 8b & DIS 2 & 150 & Chabrier \\
model 8c & DIS 2 & 350 & Chabrier \\
model 9a & DIS 3 & 0 & Chabrier \\
model 9b & DIS 3 & 150 & Chabrier \\
model 9c & DIS 3 & 350 & Chabrier \\
\hline

\hline
\end{tabular}%
\flushleft
\footnotesize{{\bf Notes.} The columns respectively provide the name of the model, the initial rotational velocity of massive stars, the duration of the pause in the SF, and the adopted IMF. The first three models are the Matteucci-like ones. We adopted the same prescriptions as the best model from \citet{Matteucci2019}, changing only the yields for massive stars. 
For the remaining models, we adopted either a constant rotational velocity for massive stars or a distribution of rotational velocities (presented in Section \ref{sec: the model}). Different durations ($\mathrm{\delta_{SF}}$) for the quenching of the SF were also tested.}
\end{table}

\subsection{Metallicity distribution function}

\begin{figure}
    \centering
    \includegraphics[width=1\columnwidth]{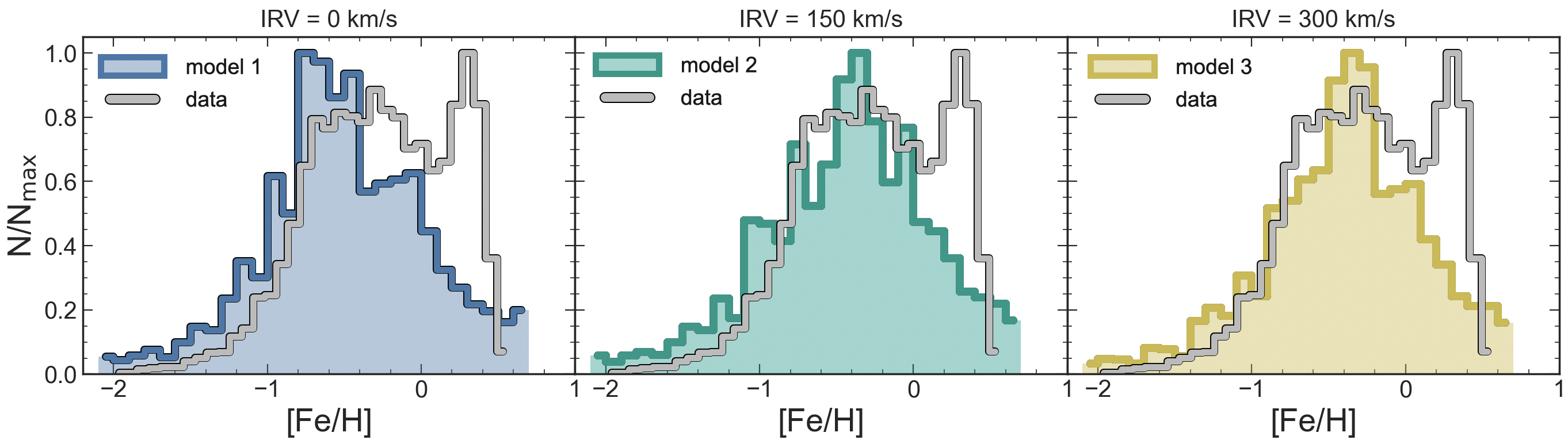}
%    \captionsetup{width=1\textwidth}
    \caption{MDFs as predicted by models 1, 2, and 3 obtained with the prescriptions of the best \citet{Matteucci2019} model with yields of rotating massive stars with initial velocities equal to 0 km/s (left panel), 150 km/s (middle panel), and 300 km/s (right panel).}
    \label{fig: Salpeter M19}
\end{figure}

In Figure \ref{fig: Salpeter M19} we report the results of model 1, model 2, and model 3 obtained with the prescriptions of \citet{Matteucci2019}'s best model but with updated yields for rotating massive stars (see Table \ref{tab: models_bulge} for reference). None of the three models is able to reproduce the MDF observed in the new set of observational data. The predicted MDFs are shifted toward low metallicities, in particular in the case of model M19-0 where the MDF is peaked at $\mathrm{[Fe/H]\simeq-0.6\ dex}$. In the case of models 2 and 3 the peak is at higher metallicities, at $\mathrm{[Fe/H]\simeq-0.3\ dex}$, since rotation in massive stars increases the production of Fe. However, the observed MR peak is still not reproduced, and, moreover, all of the models overestimate the stars in the low metallicity regime. 

\begin{figure*}
    \centering
    \includegraphics[width=1\textwidth]{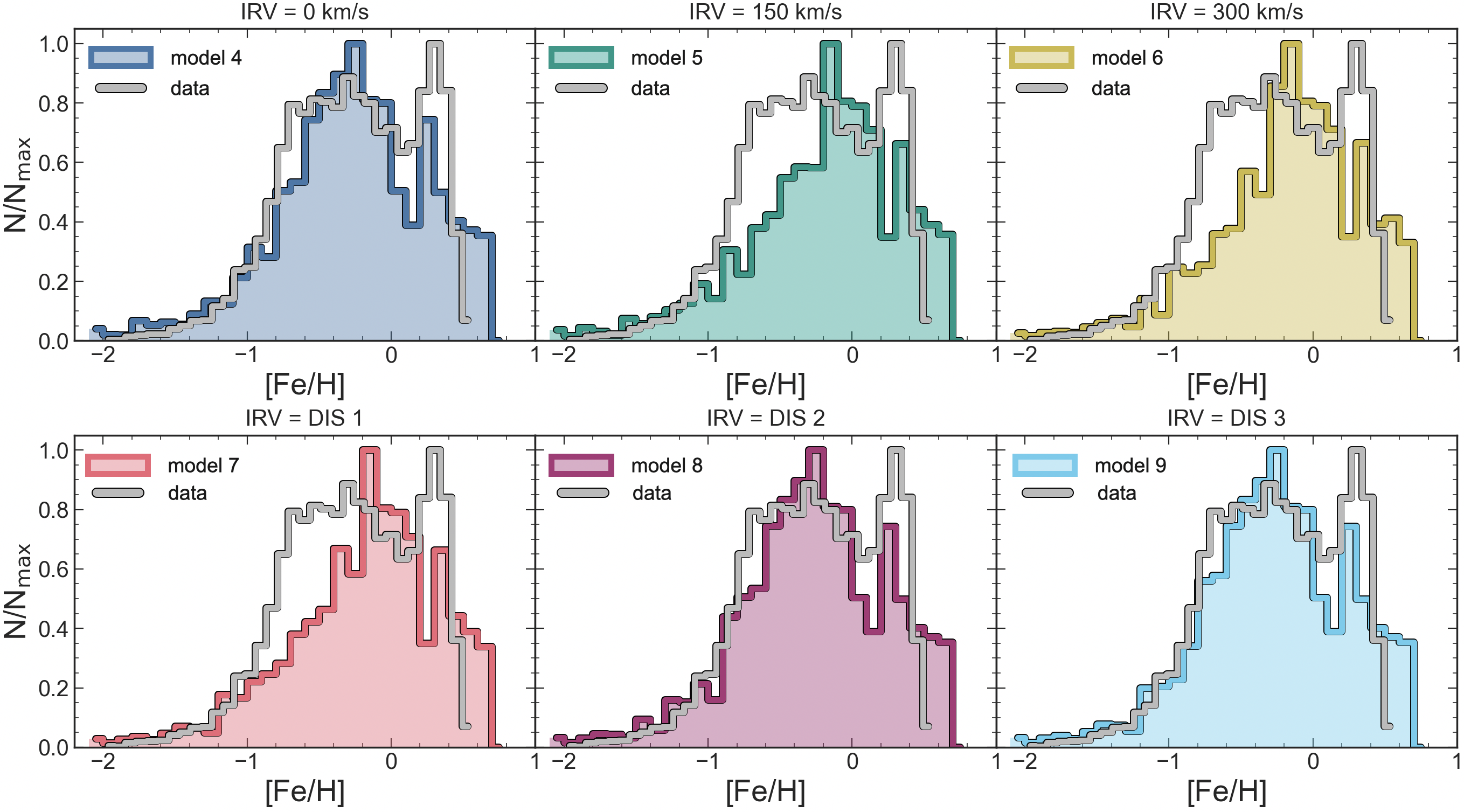}
%    \captionsetup{width=1\pagewidth}
    \caption{MDFs predicted with a Chabrier IMF and a pause in the SF lasting 250 Myr. The MDFs differ for the different initial rotational velocities adopted. Models 4, 5, and 6 (first row) have constant initial rotational velocities, and models 7, 8 and 9 are characterized by a distribution of the initial rotational velocities  (see Table \ref{tab: models_bulge}).}
    \label{fig: Bulge stop 250}
\end{figure*}

In Figure \ref{fig: Bulge stop 250} we present the results of models 4, 5, and 6 in the first row, and of models 7, 8, and 9 in the second row. All the models show an improved agreement with data with respect to old models. Specifically, models 4, 8, and 9 effectively reproduce the positions of the first and second MDF peaks. In contrast, models 5, 6, and 7 show a less good fit with the data, yielding to similar MDF shapes: the MP peak is shifted toward too high metallicities, with a notable underproduction of stars in the intermediate metallicity range ($\mathrm{-1\le[Fe/H]\le-0.25\ dex}$).

\begin{figure*}
    \centering
    \includegraphics[width=1\textwidth]{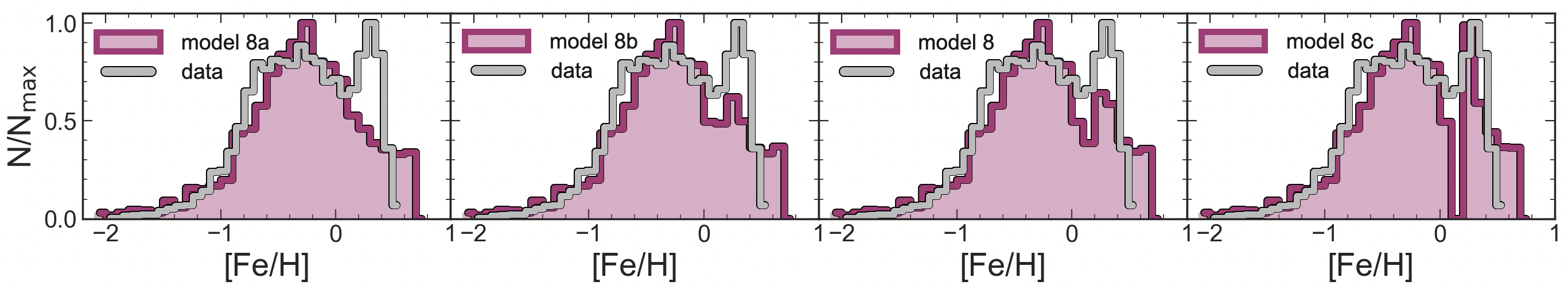}
%    \captionsetup{width=1\textwidth}
    \caption{MDFs predicted by models 8(A-C) characterized by different durations of the SF pause (0, 150, 250, and 350 Myr).}
    \label{fig: Bulge different stop}
\end{figure*}

\begin{figure*}
    \centering
    \includegraphics[width=0.8\textwidth]{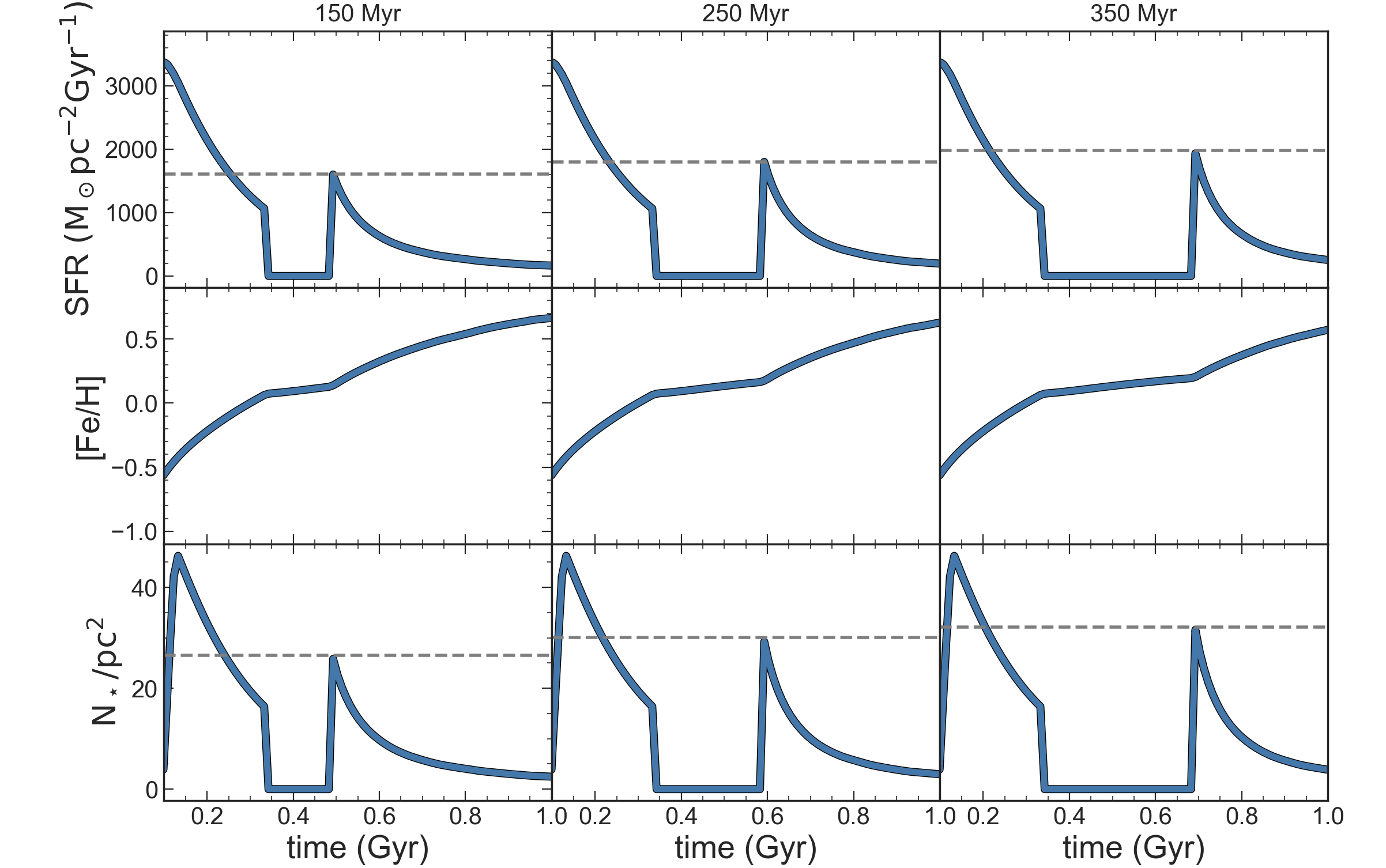}
%    \captionsetup{width=1\textwidth}
    \caption{Star formation rate, [Fe/H], and surface number density of stars as a function of time predicted by models 4, 4b, and 4c. The three models differ only in the duration of the SF quenching period.}
    \label{fig: SF_FeH_stars}
\end{figure*}

The drop produced by all of the models between the MP and the MR peak appears to be slightly too deep with respect to the observed one. The depth of the drop in the MDF increases with the duration of the SF pause. This can be seen in Figure \ref{fig: Bulge different stop}, where we increase the duration of the SF pause from 0 to 150, 250, and 350 Myr (for illustration purposes we show results only for models 8(A-C)). The longer the SF pause lasts, the deeper the drop will get in the MDF and, as a consequence, a higher second peak in the MDF will be obtained. In the time interval during which  the SF is quenched, no new SF occurs and the production of Fe slows down because of the lack of new massive stars (Type Ia SNe are still active, and therefore the production of Fe has not completely stopped). While the SF is being quenched, the gas builds up due to the ongoing infall; hence, more gas has accumulated by the end of a longer pause. When the SF resumes, since it is proportional to the gas available in the ISM, it will be stronger after a longer pause, and as a consequence the stellar number density will be larger. This can be seen in Figure \ref{fig: SF_FeH_stars}, where we report the evolution of the SF, [Fe/H], and surface number density of stars as a function of time for pauses in the SF of 150, 250, and 350 Myr. Thus, the height of the MR peak can be modeled by changing the duration of the SF pause or, as recently analyzed by \citet{Romano2023} for the chemical enrichment of the bulge fossil Terzan 5, by removal of a major fraction of the gas left over from the first SF episode.

\subsection{Abundance patterns}

\begin{figure*}
\begin{center}
    \subfloat{\includegraphics[width=1\textwidth]{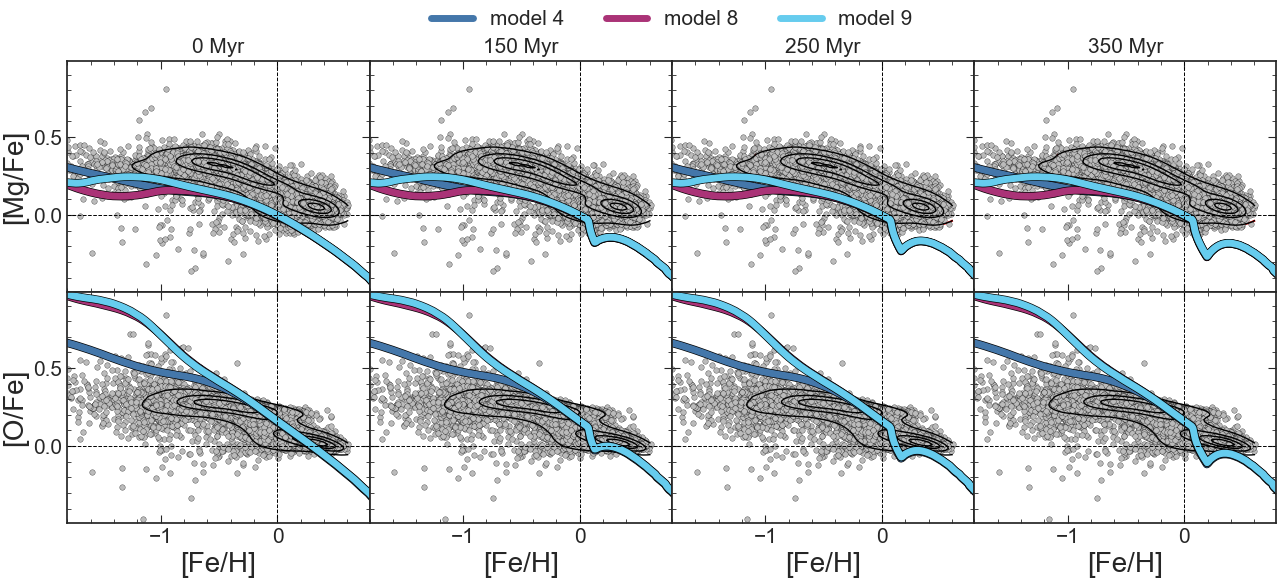}}
    \vfill
    \subfloat{\includegraphics[width=1\textwidth]{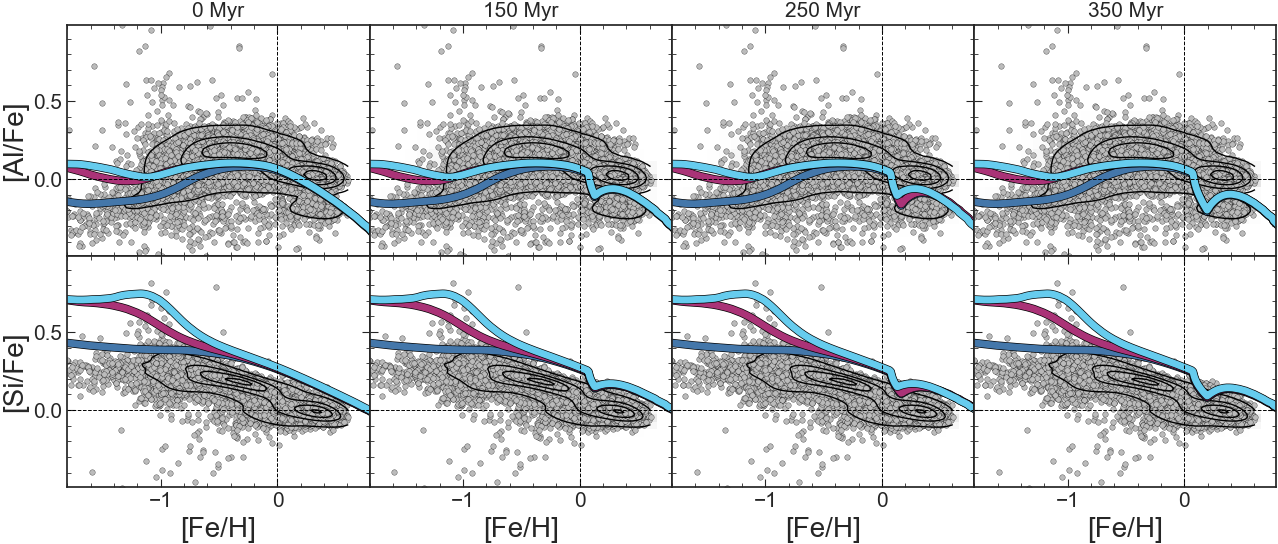}}
    \caption{Predicted [Mg/Fe], [O/Fe], [Al/Fe], and [Si/Fe] vs. [Fe/H] trends for models 4(a-c), 8(a-c), and 9(a-c) without the contribution from accreted disk stars. The models differ in the prescriptions on the initial rotational velocities of massive stars and the duration of the SF quenching period (see Table \ref{tab: models_bulge}).}
    \label{fig: Mg O Al Si}
\end{center}
\end{figure*}

In Figure \ref{fig: Mg O Al Si} we report the [$\alpha$/Fe] versus [Fe/H] abundance patterns of Mg, O, Al, and Si as predicted by models 4(a-c), 8(a-c), and 9(a-c). Models with a pause in the SF produce a break in the [$\alpha$/Fe] versus [Fe/H] that is also visible in the APOGEE data, which show indeed two overdense regions in correspondence of [Fe/H]$\simeq$-0.5 dex and [Fe/H]$\simeq$0.25 dex. The break in the abundance patterns occurs because the pause in the SF causes cessation of $\alpha$-element production in massive stars, while the Fe production continues (even if it is significantly slowed) thanks to Type Ia SNe and their longer time-delays. The bimodal distribution that is present in the set of data used in this work is also reported by \citet{Queiroz2021} (see also \citealp{Rojas-Arriagada2019, Queiroz2020}), although in their case the depression between the two peaks seems to be more pronounced, and the two sequences are more noticeably distinct. However, the different datasets are comparable and their differences are not affecting our conclusions.

\begin{figure}
    \centering
    \includegraphics[width=1.0\columnwidth]{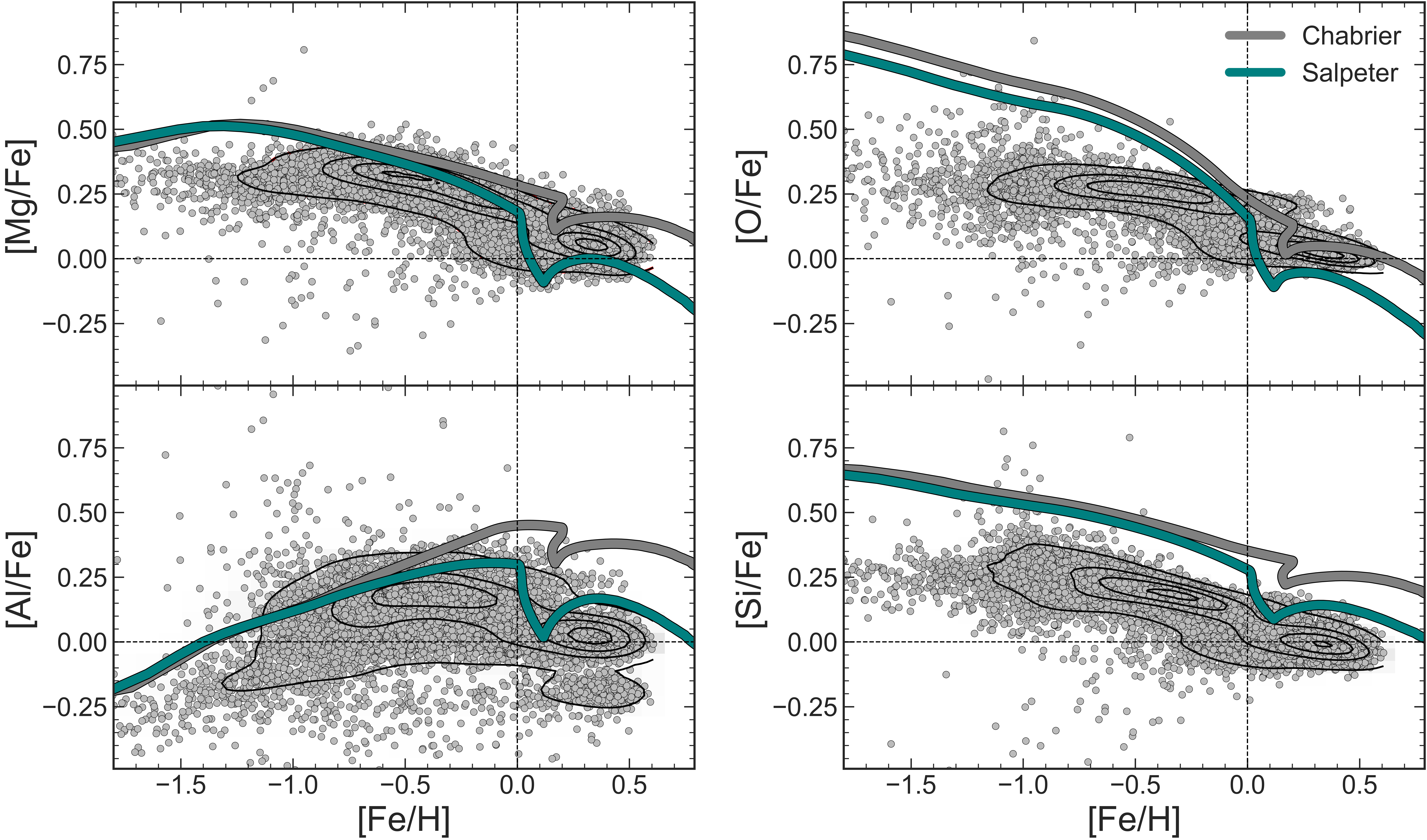}
    \caption{Predicted [Mg/Fe], [O/Fe], [Al/Fe], and [Si/Fe] vs. [Fe/H] trends for model 4 but with yields set for massive stars from \citet{Kobayashi2006} and different IMFs.}
    \label{fig: Kobayashi}
\end{figure}

\begin{figure}
    \centering
    \includegraphics[width=1.0\columnwidth]{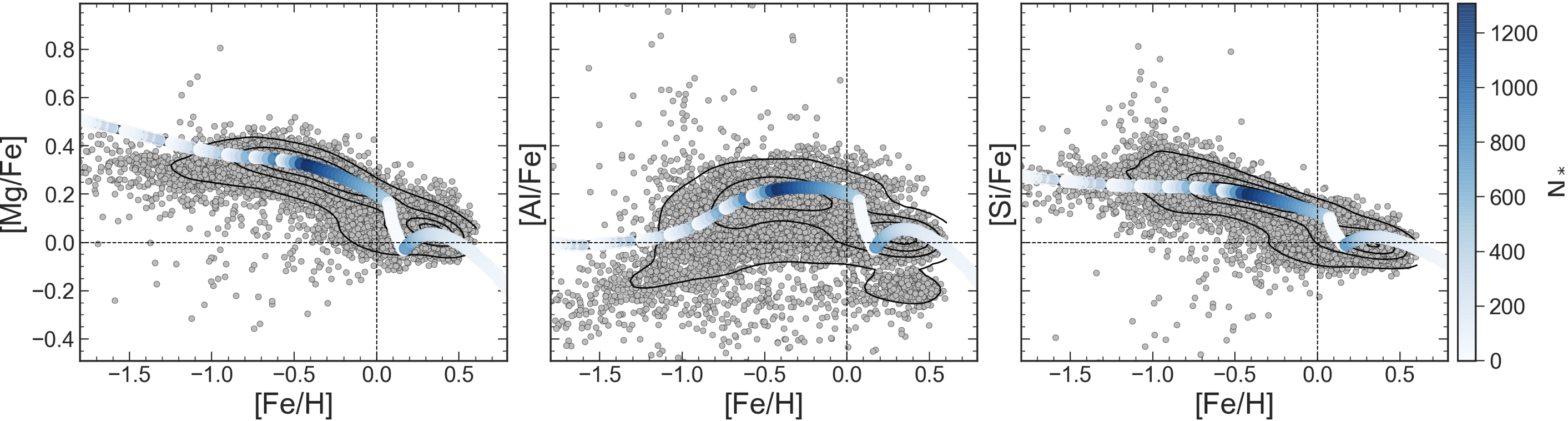}
    \caption{Predicted [Mg/Fe], [Al/Fe], and [Si/Fe] vs. [Fe/H] trends for model 4 with corrected yields for massive stars.}
    \label{fig: Mg_Al_Si_corrected}
\end{figure}

Rotation in massive stars increases the production of O, Al, and Si, while it reduces that of Mg. Since the theoretical distributions of rotational velocities differ mainly at low metallicities, this effect is visible only for [Fe/H]$\le$-1 dex. All the models are underestimating the [Mg-Al/Fe] versus [Fe/H] trends and overestimating the [Si/Fe] one. Models C-0(A-D) are those that best reproduce the [Al/Fe] versus [Fe/H] as well as the [O/Fe] versus [Fe/H], even if deviations at low metallicities are present. The general trend of the observed abundance ratios is, however, well reproduced by models 4(a-c), in particular with a pause of $\mathrm{250\ Myr}$. Deviations from the observed pattern are therefore due to the adopted set of yields, the IMF, or a combination of the two. A \citet{Salpeter1955} IMF would indeed predict lower abundance patterns for the $\alpha$ elements, because of a lower production of massive stars. Therefore, by keeping the same set of yields, this would improve the agreement with the [O/Fe] and  [Si/Fe] versus [Fe/H] trends but, at the same time, would worsen the ones with the [Mg/Fe] and [Al/Fe] as well as with the MDF (as shown in the previous section). Concerning the yields variations, adopting the set of \citet[which accounts for mass loss but not rotation]{Kobayashi2006} produces the results shown in Figure \ref{fig: Kobayashi} for both a \citet{Salpeter1955} and a \citet{Chabrier2003} IMF. The [Mg/Fe] versus [Fe/H] trend is improved with respect to model 4, in particular by adopting a \citet{Salpeter1955} IMF (see \citealp{Matteucci2019}). However, O, Al and Si the fit with the data is poorly improved or gets even worse. If the assumption of a \citet{Chabrier2003} is valid, then corrections to the \citet{Limongi2018} yields must be applied. According to our model, the production of Mg and Al should be increased by a factor of $\sim1.5$, while that of Si should be decreased by a factor of $\sim0.7$. When those corrections are applied, the derived results for the [Mg-Al-Si/Fe] abundance patterns are those shown in Figure \ref{fig: Mg_Al_Si_corrected} as a function of both the [Fe/H] and the number of stars produced. In this way, it is possible to better appreciate the decrease in the stellar density that is produced by our model at $\mathrm{0.0\leq[Fe/H]\leq0.2}$, compatible with the observed data.

\begin{figure}
\begin{center}
 \subfloat{\includegraphics[width=1\columnwidth]{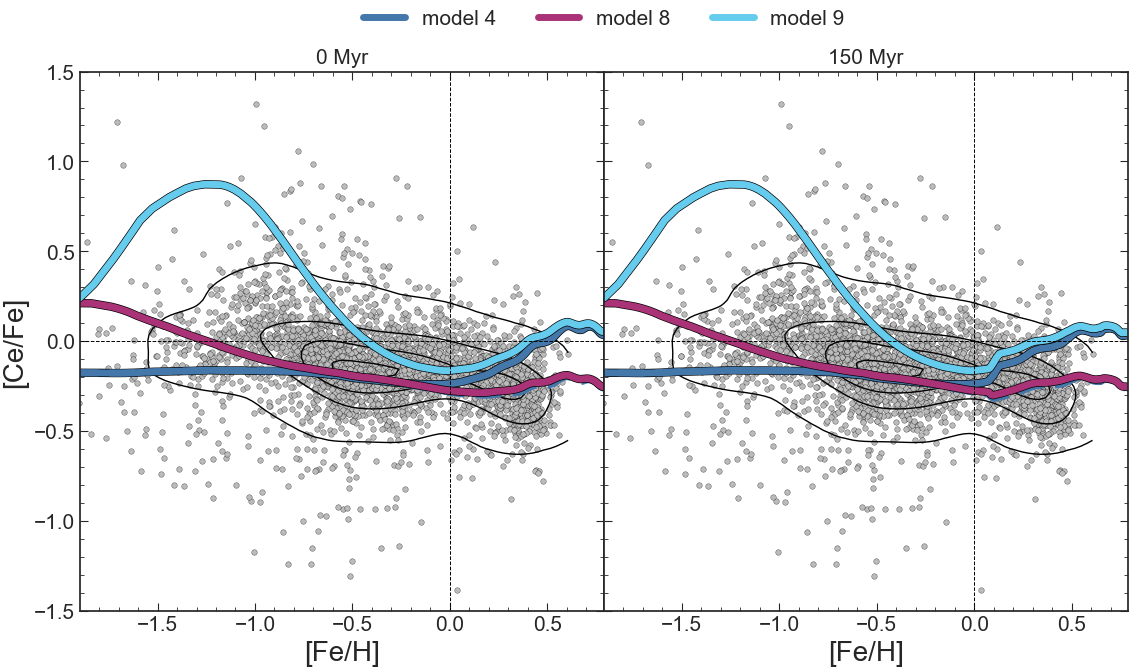}}
\vfill
 \subfloat{\includegraphics[width=1\columnwidth]{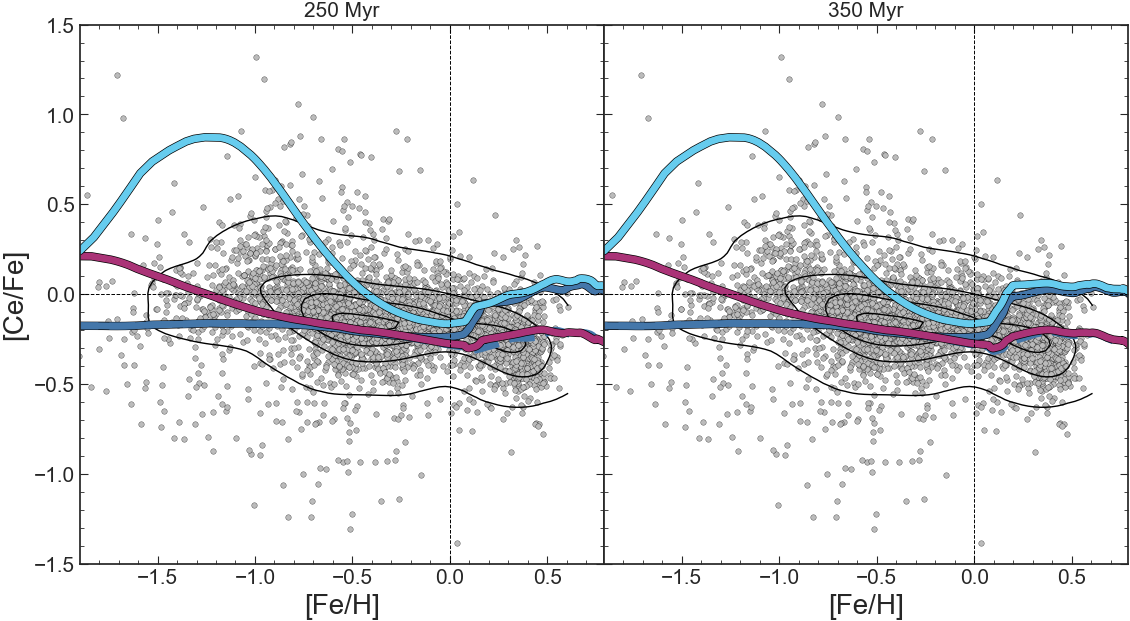}}
 \caption{Same as Figure \ref{fig: Mg O Al Si} but for Ce. The dashed purple line corresponds to model 8 but with reduced yields from LIMSs.}%
 \label{fig: CeFe}%
\end{center}
\end{figure}

In Figure \ref{fig: CeFe} we report results of models 4(a-c), 8(a-c) and 9(a-c) for the [Ce/Fe] versus [Fe/H] abundance pattern. The characteristic ``banana'' shape of s-process elements can be seen in the data at $\mathrm{-0.77\leq [Fe/H]\leq 0.60\ dex}$. At lower [Fe/H], dispersion in the data is also present with mean value $\mathrm{[Ce/Fe]\simeq0.20\ dex}$. The bimodal distribution clearly seen in the $\alpha$ elements is less evident for Ce. This can be due both to the larger scatter observed in the Ce abundance pattern and, most importantly, to the fact that Ce is also produced by LIMSs (its s-process component) and by MNSs (its r-process component), and therefore its production will continue even if a quenching in the SF is present. Indeed, in the model results as well, the different duration of the SF pause has very little impact on the overall [Ce/Fe] versus [Fe/H] pattern, without affecting too much the comparison with the data. For the s-process elements, such as Ce, rotation in massive stars increases their production, especially at low metallicities. As already discussed in \citet{Molero2023}, it is possible to note the strong effect of the magneto-rotational SNe at low metallicities responsible for the creation of the plateau visible in models 4(a-c). When higher rotational velocities for massive stars are considered, their contribution dominates at low [Fe/H]. However, because of such high velocities, models 9(a-c) are overproducing the [Ce/Fe] versus [Fe/H] at low [Fe/H], while models 4(a-c) and 8(a-c) are more in agreement with the data. At higher [Fe/H], LIMSs dominate the production of s-process elements, creating a bump in the [Ce/Fe]. With the adopted sets of yields, however, the production of Ce from LIMSs is too strong, and it should be reduced of at least a factor of 3 in order to have a better agreement with the data, as shown by the dashed purple line in the figure, corresponding to models 4(a-c), but with reduced AGB yields. 

\section{Results with accreted stars}
\label{sec: ch6 results with the bar} 

As discussed in Section \ref{sec: Introduction}, in this work we are testing the hypothesis put forward by \citet{Matteucci2019}, according to which the MR peak of the Galactic bulge is due to both stars that are formed in situ and to stars that formed in the innermost part of the Galactic disk and got trapped in the Galactic bar. This is also supported by the MDF of the innermost part of the Galactic disk obtained by our model, which presents a peak at a similar value of [Fe/H] to that for the MR population in the bulge. This is shown in Figure 3 of \citet{Matteucci2019}, where the adopted model for the disk is the one-infall model of \citet{Grisoni2017} with a constant SF efficiency of $1\ \rm Gyr^{-1}$, a \citet{Kroupa1993} IMF and nucleosynthesis prescriptions from \citet{Romano2010}. Here, for consistency with our previous more recent works, we used the delayed two-infall model presented by \citet{Palla2020} and adopted by \citet{Molero2023} to model the evolution of several neutron-capture elements in the MW disk (see also \citealp{Spitoni2019, Spitoni2021}). At variance with the \citet{Grisoni2017} model, we adopted a higher SF efficiency of $9\ \rm Gyr^{-1}$ (as opposed to $\mathrm{1\ Gyr^{-1}}$)  in the inner region of the disk, as suggested by \citet{colavitti2009} (see also \citealp{Spitoni2015}). The nucleosynthesis prescriptions are from \citet{Molero2023} best model and a \citet{Kroupa1993} IMF is adopted. 

\subsection{Metallicity distribution function}

\begin{figure*}
\begin{center}
 \subfloat{\includegraphics[width=1\textwidth]{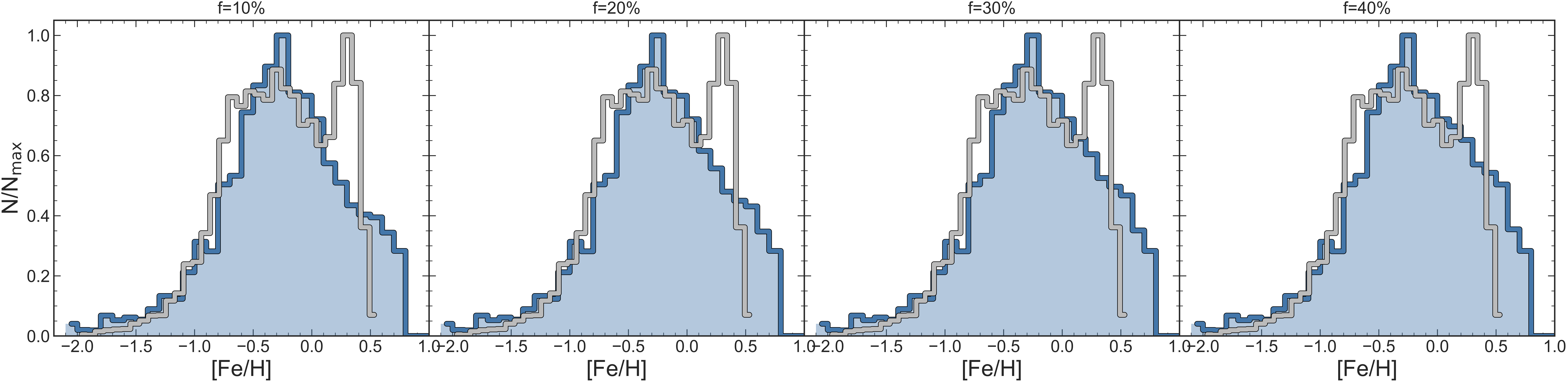}}
\vfill
 \subfloat{\includegraphics[width=1\textwidth]{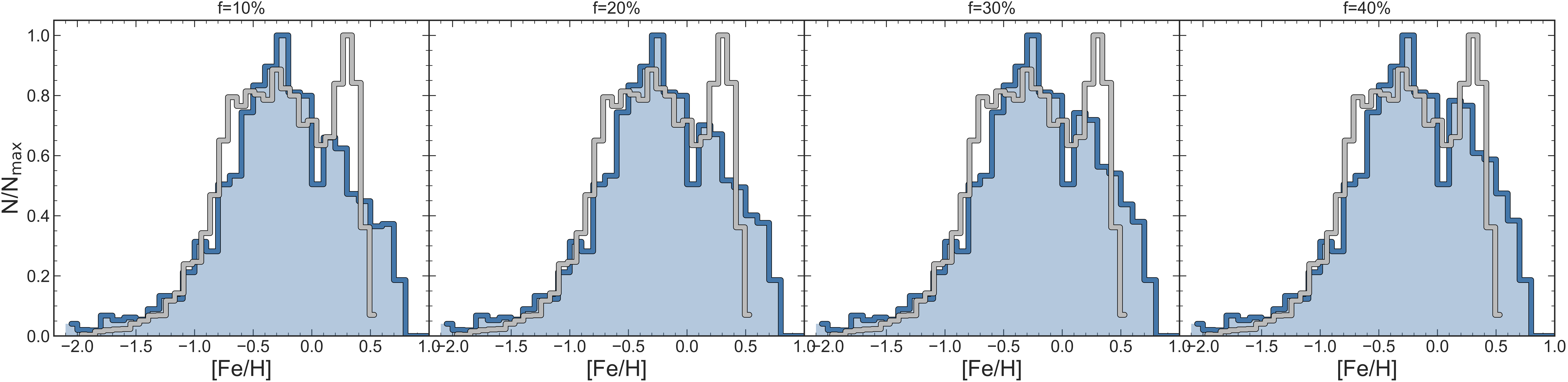}}
\vfill
 \subfloat{\includegraphics[width=1\textwidth]{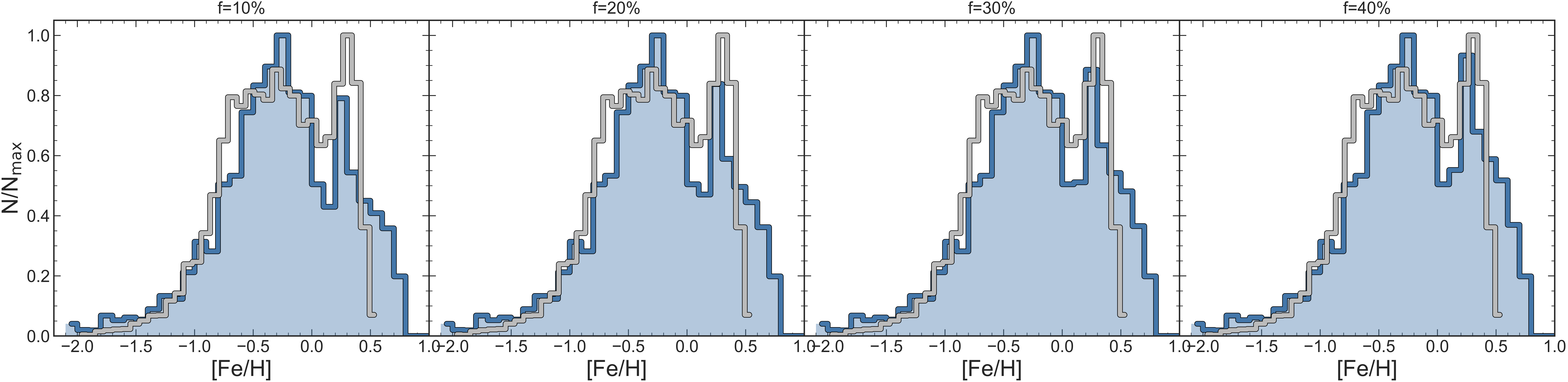}}
 \caption{MDF of the Galactic bulge predicted by models 4, 4a, and 4b with no pause in the SF (first row), 150 Myr of pause (second row), and 250 Myr (third row, model 4) and with different fractions of inner disk stars contributing to the bar.}%
 \label{fig: MDF bar}%
\end{center}
\end{figure*}

\begin{figure}
    \centering
    \includegraphics[width=0.8\columnwidth]{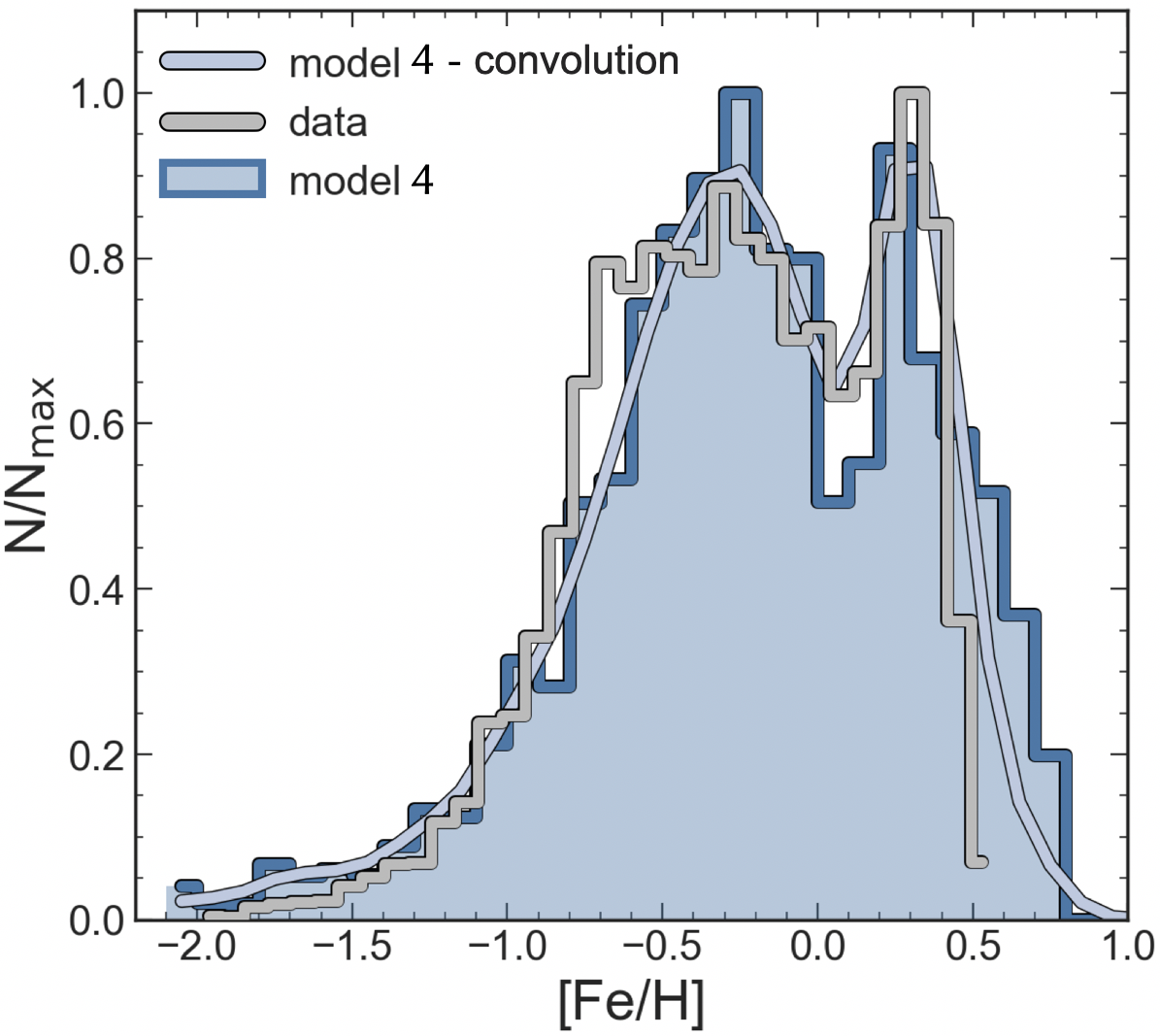}
    \caption{MDF of the Galactic bulge as predicted by model 4 with a fraction of $\mathrm{40\%}$ of stars from the inner disk. The model has been convoluted with a normal distribution with $\mathrm{\sigma=0.1\ dex}$ to conservatively take the spectroscopic uncertainly in metallicity into account.}
    \label{fig: convolution}
\end{figure}

In Figure \ref{fig: MDF bar} we present the results of models 4(a-c) for the MDF, with the contribution of stars from the inner disk. To account for the fraction of stars, initially calculated as an upper limit in Section \ref{sec: the model}, we performed tests with different values, namely $\mathrm{f=10,\ 20,\ 30,\ 40\%}$. As shown in the first row of the figure, when we do not consider any pause in SF, the inclusion of a certain percentage of stars from the Galactic disk alone does not suffice to replicate the MR peak observed in the MDF. However, by assuming a SF pause of $\mathrm{150\ Myr}$ leads to an improved agreement with the data. Model 4, with a $\mathrm{250\ Myr}$ SF pause and $\mathrm{f=40\%}$ stars from the inner disk, best reproduces the observed results. Both the MP and the MR peaks within the MDF are notably well matched, not only in terms of their positions but also in terms of their relative heights. The model, however, still produces a slightly deeper hole between the two peaks, as well as an overproduction of star for [Fe/H]$\ge 0.5\ \mathrm{dex}$. This two issues are nicely solved by performing a convolution between the distribution predicted by our model 4 with a normal distribution with $\mathrm{\sigma=0.1\ dex}$. In fact, as shown in Figure \ref{fig: convolution}, the two discrepancies are well inside the uncertainties of the model and data.

\subsection{Abundance patterns}

\begin{figure*}
    \centering
    \includegraphics[width=1.0\textwidth]{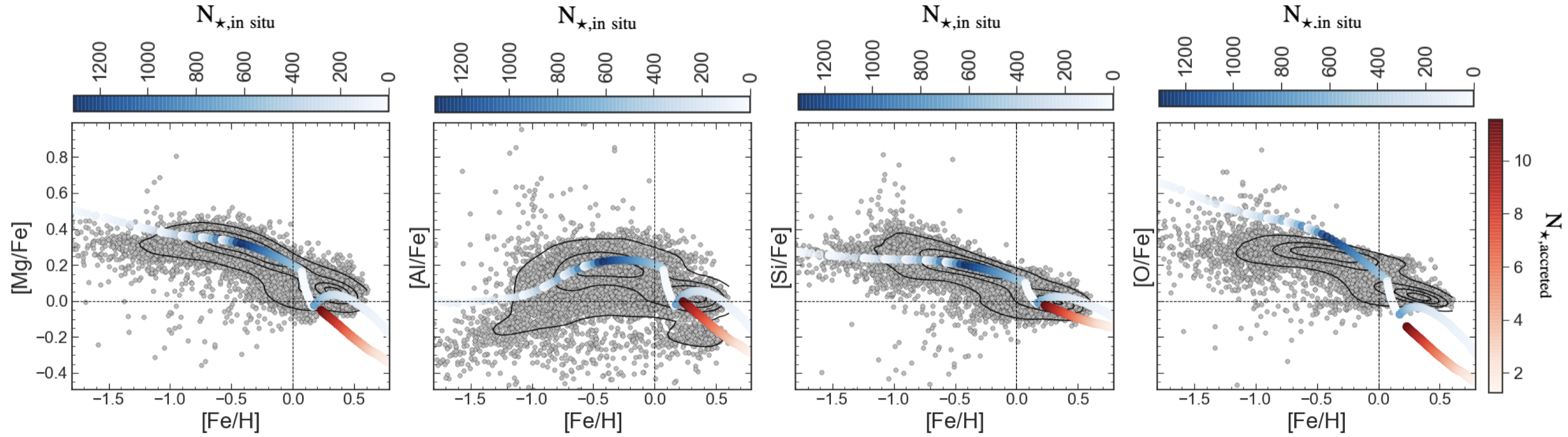}
    \caption{[$\alpha$/Fe] versus [Fe/H] abundance patterns for Mg, Al, Si, and O as predicted by model 4 (upper blue lines) responsible for the creation of the in situ stellar population and by stars from the inner disk (lower red lines) responsible for the creation of the accreted stellar population in the bulge. Lines are color-coded by the number of stars formed by the model, with the upper blue colorbar referring to the in situ population and the right red colorbar referring to the accreted population.}
    \label{fig: abundances_bulgebar}
\end{figure*}

In Figure \ref{fig: abundances_bulgebar} we report results for the [$\alpha$/Fe] versus [Fe/H] abundance patterns. The blue lines in the diagrams refer to the chemical enrichment of stars born in situ in the bulge, while the red lines refer to stars born in the inner disk of the MW and then accreted by the growing bar. The contribution from the accreted population appears in the [$\alpha$/Fe] versus [Fe/H] diagrams from [Fe/H]$\mathrm{\simeq+0.2\ dex}$, which corresponds to the value of metallicity at which the bulge starts growing the bar. This value of metallicity is in agreement with the position of the MR cluster of data in the $\alpha$-element abundance patterns and is obtained by assuming that the bar age is $\mathrm{\sim8\ Gyr}$, namely that the accretion of stars from the inner disk started $\mathrm{\sim 5.5\ Gyr}$ ago. The stars being accreted will enrich the ISM in the bulge region with the product of their chemical composition, which will reflect that of the location where they were trapped. Therefore, according to the time-delay model (see \citealp{Matteucci2012}), we expect their abundance patterns to lie below that due to the stellar population born in situ in the bulge, as shown in the figure. The abundance patterns predicted by our model for the accreted population is well in agreement with the bulge MR data for [Al/Fe] and for [Si/Fe] versus [Fe/H], but underproduce data for [Mg/Fe] and for [O/Fe] versus [Fe/H]. This can potentially be due to the adopted IMF for modeling the inner MW disk, here namely a \citet{Kroupa1993} IMF. Among the different IMFs, the \citet{Kroupa1993} IMF is the one that guarantees good fits to the observed properties of the solar neighborhood (see \cite{Romano2005}). In chemical evolution studies, this parametrization for the IMF is commonly adopted not only for the solar vicinity, but also for modeling the all MW disk. However, there is currently no strong observational evidence that the \citet{Kroupa1993} IMF provides the best fit with the inner MW disk as well. It is reasonable to presume that, moving outward in radius from the Galactic bulge toward the inner disk and the solar neighborhood, the slope of the IMF may get steeper. So that, if a \citet{Chabrier2003} IMF can be adopted for the Galactic bulge (as shown in the previous sections) and a \citet{Kroupa1993} is ideal for the MW disk around $\mathrm{R_{GC}\sim8\ kpc}$, then a \citet{Salpeter1955} IMF may be the right one to model the innermost part of the MW disk.

\section{Discussion and conclusions}

Our model of the Galactic bulge is based on the model from \citet{Matteucci2019}, and it is mainly constrained to the shape of the bulge MDF as observed in the sample of $\sim$13000 giant stars from APOGEE DR16 (\citealp{Rojas-Arriagada2020}). Our main assumption is that the MW bulge was formed by a strong SF episode, triggered by a fast collapse of primordial gas, which is mainly responsible for the creation of the MP population observed in the MDF as well as in the [$\alpha$/Fe] versus [Fe/H] abundance patterns. A second, smaller burst of SF and an accretion of disk stars from the Galactic bar are assumed to be responsible for the creation of the MR stellar population. Our main conclusions are the following:

\begin{itemize}
    \item The best \citet{Matteucci2019} model, which assumes a delay between the two SF bursts of $\mathrm{250\ Myr}$, is not able to reproduce the observed MDF of the bulge represented by the \citet{Rojas-Arriagada2020} dataset if yields for rotating massive stars from \citet{Limongi2018} are used together with a \citet{Salpeter1955} IMF.
    \item By adopting a \citet{Chabrier2003} IMF, rather than a \citet{Salpeter1955} one, the agreement between the \citet{Matteucci2019} model and the data is improved. In particular, a model that assumes no rotational velocity for massive stars or a distribution of rotational velocities that favors slow rotation at the present time best reproduces the observed MDF.
    \item The observed abundance trends of $\alpha$ elements are well reproduced with a delay between the two SF bursts of $\mathrm{250\ Myr}$, but all the models underproduce the [Mg/Fe] and [Al/Fe] versus [Fe/H] relations and overproduce the [Si/Fe] one. The situation is slightly improved by using a different set of yields (the one from \citealp{Kobayashi2006}), but only in the case of Mg. We conclude that, if the assumption of a \citet{Chabrier2003} IMF is valid, then corrections need to be applied to the \citet{Limongi2018} set of  yields (R). The production of Mg and Al should be increased by a factor of $\mathrm{\sim1.5}$, while that of Si should be decreased by a factor of $\mathrm{\sim 0.7}$.  
    \item A model with no massive star rotation also best reproduces the [Ce/Fe] versus [Fe/H] trend, mainly because of the contribution from magneto-rotational SNe at low metallicities. However, the abundance ratios are overestimated at high [Fe/H] because of the overly strong impact of AGB stars, whose contribution to s-process element nucleosynthesis, in agreement with other chemical evolution studies (e.g., \citealp{Rizzuti2019, Rizzuti2021, Molero2023}), should be decreased by at least a factor of 2.
    \item To reproduce the MR peak of the MDF, we estimate that at least $\sim40\%$ of stars should get trapped by the growing bar. However, if no pause in the SF is assumed, contamination of stars from the inner disk is not sufficient to fully reproduce the MR peak of the MDF; therefore, a pause in the SF of $\mathrm{250\ Myr}$ must be considered. 
    \item The accreted population from the innermost part of the disk is able to fit the MR data relative to the [$\alpha$/Fe] versus [Fe/H] diagrams in the case of Al and Si, but it underproduces the abundances for Mg and O. We suggest that this can be reconciled by adopting a steeper IMF.
\end{itemize}

In summary, to fit the bimodal shape of the bulge MDF, one of the main assumptions of our model is a two-phase SF history consisting of an initial starburst followed by a quenching episode and a second, less intense burst. During the first burst, the MP and high-$\alpha$ bulge stars are formed. In particular, the rapid SF  cessation process appears critical for reproducing the observed alpha bimodality as well as the well-defined depression between the two MDF peaks (see also \citealp{Lian2020, Lian2021}). Simulations of MW mass halos suggest that the gas infall rate peaks around redshift $\mathrm{z = 2}$ and then declines (see, e.g., \citealp{vandevoort2011,woods2014}). After the peak, the MW inevitably starts quenching its SF (see also \citealp{Haywood2016}). According to \citealp{Haywood2018}, even if the decline in the gas infall rate coincides with the quenching in the SF, it is not necessarily its main cause. In fact, in their model, there is still gas in the MW disk during this epoch, as has been observed in MW-type galaxies at redshifts around z = 1.5 (\citealp{daddi2010}). As a consequence, even if a decline in the infall rate might have contributed to the SF quenching, it is not necessarily its main (or unique) cause. N-body hydrodynamic simulations from \citet{Khoperskov2018} show that it is possible for the bar to suppress the SF. In their simulations, they quantitatively explore the relation between bar formation and SF in gas-rich galaxies and conclude that the action of a stellar bar efficiently quenches the SF, reducing its rate by a factor of 10 in less than 1 Gyr. The SF efficiency decreases and, in all of their models, the bar quenches the SF in the galaxy. Similarly, \citet{Haywood2016} propose that in the MW the action of the bar could have increased the gas turbulence, stabilizing the disk against SF and quenching its SF. As a consequence, the suppression of the SF is not associated with a substantial consumption of the available gas. The gas would still be present in the disk, but its high turbulence would significantly limit the SF efficiency. However, it must be noted that in order to fit the two peaks in the observed MDF, the quenching in the SF would have to happen at earlier times with respect to the formation of the bar. As a consequence, the most plausible explanation would be that the first, strong burst of SF consumed the majority of the available gas, causing the quenching. However, since during the quenching phase the infall of primordial gas is still active, the gas is accumulated until it triggers a second, less intense burst of SF.

We also stress that our model does not take the contribution from stars from the halo into account, nor accreted globular cluster debris or dwarf galaxies (see, e.g., \citealp{fernandez2021}). However, conclusions from our work do not necessarily exclude contributions from other astrophysical environments to the chemical enrichment of the Galactic bulge. Indeed, as underlined in the Introduction, the Galactic bulge formation mechanism is most probably a combination of different processes. These processes might include not only strong gas accretion together with secular processes, but also major and/or minor mergers. We are inclined to think that possible contributions from small satellite galaxies and/or part of the stellar halo would not change our general conclusions. In support of this hypothesis, \citet{Portail2017} conclude that only part of the very MP region of the MDF is due to the inner part of the stellar halo. If that is the case, this would influence the shape of the first peak of the MDF predicted by our model, which would suggest a less strong first SF episode; however, our conclusions on the fraction of stars that are trapped by the bar and yields from massive stars would remain unaffected. Moreover, the [$\alpha$/Fe] and the metallicity content of those stars are generally lower than the bulk of bulge stars, and as a consequence, their influence on the chemical evolution should be minimal. In particular, according to the Chemical Origins of Metal-poor Bulge Stars (COMBS), the MP bulge has higher levels of Ca and a lower dispersion in the $\alpha$-element abundances compared to halo stars of the same metallicity (see \citealp{Lucey2019}). Similarly, high neutron-capture abundance stars are found at a much lower rate in the bulge than in the halo or in MP dwarf spheroidal galaxies, and their abundance peculiarities in neutron-capture elements can often be justified with combinations of several nucleosynthetic processes (see \citealp{Koch2016, Koch2019}).

\begin{acknowledgements}
M. Molero thanks Rimpei Chiba and Mattia Sormani for the very useful discussions and suggestions concerning bar dynamics. M. Molero thanks Marco Limongi for interesting discussions concerning stellar rotation. F. Matteucci thanks I.N.A.F. for the 1.05.12.06.05 Theory Grant - Galactic archaeology with radioactive and stable nuclei. A. R. A. acknowledges support from DICYT through grant 062319RA. This research was partially supported by the Munich Institute for Astro-, Particle and BioPhysics (MIAPbP) which is funded by the Deutsche Forschungsgemeinschaft (DFG, German Research Foundation) under Germany´s Excellence Strategy – EXC-2094 – 390783311. This work was supported by the Deutsche Forschungsgemeinschaft
(DFG, German Research Foundation) – Project-ID 279384907 – SFB 1245, the State of Hessen within the Research Cluster ELEMENTS (Project ID 500/10.006).
\end{acknowledgements}

\bibliographystyle{aa} % style aa.bst
\bibliography{Bibliography} % your references Yourfile.bib

%\begin{thebibliography}{}
%
%  \bibitem[Baker(1966)]{baker} Baker, N. 1966,
%      in Stellar Evolution,
%      ed.\ R. F. Stein,\& A. G. W. Cameron
%      (Plenum, New York) 333
%
%   \bibitem[Balluch(1988)]{balluch} Balluch, M. 1988,
%      A\&A, 200, 58
%
%   \bibitem[Cox(1980)]{cox} Cox, J. P. 1980,
%      Theory of Stellar Pulsation
%      (Princeton University Press, Princeton) 165
%
%   \bibitem[Cox(1969)]{cox69} Cox, A. N.,\& Stewart, J. N. 1969,
%      Academia Nauk, Scientific Information 15, 1
%
%   \bibitem[Mizuno(1980)]{mizuno} Mizuno H. 1980,
%      Prog. Theor. Phys., 64, 544
%   
%   \bibitem[Tscharnuter(1987)]{tscharnuter} Tscharnuter W. M. 1987,
%      A\&A, 188, 55
%  
%   \bibitem[Terlevich(1992)]{terlevich} Terlevich, R. 1992, in ASP Conf. Ser. 31, 
%      Relationships between Active Galactic Nuclei and Starburst Galaxies, 
%      ed. A. V. Filippenko, 13
%
%   \bibitem[Yorke(1980a)]{yorke80a} Yorke, H. W. 1980a,
%      A\&A, 86, 286
%
%   \bibitem[Zheng(1997)]{zheng} Zheng, W., Davidsen, A. F., Tytler, D. \& Kriss, G. A.
%      1997, preprint
%\end{thebibliography}

\end{document}